\documentclass[a4paper,fleqn,usenatbib]{mnras2}
\usepackage{newtxtext,newtxmath}
\usepackage[T1]{fontenc}
\usepackage{ae,aecompl}
\usepackage{graphicx}	% Including figure files
\usepackage{amsmath}	% Advanced maths commands
\usepackage{amssymb}	% Extra maths symbols
\usepackage{natbib}
\usepackage{color}
\usepackage{ulem}
\usepackage{mathtools}
\usepackage{threeparttable}
\usepackage{booktabs}
\usepackage{verbatim}

\title[The He star donor channel towards PSR J1953+1844]{The He star donor channel towards the black widow PSR J1953+1844}

\author[Y. Guo et al.]{
	Yunlang Guo,$^{\rm 1,2,3}$\thanks{E-mail:yunlang@nju.edu.cn}
	Bo Wang$^{\rm 3,4,5}$\thanks{E-mail:wangbo@ynao.ac.cn}
	and
	Xiangdong Li$^{\rm 1,2}$\thanks{E-mail:lixd@nju.edu.cn}
	\\
	$^{1}$School of Astronomy and Space Science, Nanjing University, Nanjing 210023, China\\
	$^{2}$Key Laboratory of Modern Astronomy and Astrophysics, Nanjing University, Ministry of Education, Nanjing 210023, China\\
	$^{3}$University of Chinese Academy of Sciences, Beijing 100049, China\\
	$^{4}$Key Laboratory for the Structure and Evolution of Celestial Objects, Yunnan Observatories, Chinese Academy of Sciences, Kunming 650216, China\\
	$^{5}$International Centre of Supernovae, Yunnan Key Laboratory, Kunming 650216, China\\
}

\date{Accepted XXX. Received YYY; in original form ZZZ}

\pubyear{2023}

\begin{document}
	\label{firstpage}
	\pagerange{\pageref{firstpage}--\pageref{lastpage}}
	\maketitle
	
\begin{abstract}
Black widows (BWs) are a type of eclipsing millisecond pulsars (MSPs)
with low companion masses ($\lesssim0.05\,\rm M_\odot$) and tight orbits ($<1$\,d).
PSR J$1953+1844$ (i.e. M71E)
is a BW with the shortest orbital period ($\sim53$ minutes) ever discovered,
which was recently detected by
Five-hundred-meter Aperture Spherical  radio Telescope.
Its companion mass is $\sim0.01\,\rm M_\odot$ according to its mass function,
indicating that the companion may be a hydrogen-deficient star.
However,
the origin of PSR J$1953+1844$ is highly unclear.
In this paper,
we explored the origin of PSR J$1953+1844$
through
%by considering the evaporation process based on
the neutron star+He star channel,
in which the system can experience ultracompact X-ray binary phase.
%Meanwhile,
We found that
the He star donor channel can reproduce
the characteristics of PSR J$1953+1844$,
indicating that this work provides an alternative formation channel for this source.
Meanwhile,
the minimum orbital period of BWs
formed by this channel is $\sim28$ minutes,
corresponding to the companion mass of $0.058\,\rm M_\odot$.
In addition,
we note that even though PSR J$1953+1844$ has a short orbital period,
it cannot be detected by the gravitational wave (GW) observatories like
Laser Interferometer Space Antenna, TaiJi and TianQin.
However,
we still expect that the BWs with extremely tight orbit
produced by this channel are the potential sources
of future space-based GW observatories.
Moreover,
our simulations show that
PSR J$1953+1844$ may eventually evolve into an isolated MSP.
		
\end{abstract}
	

\section{Introduction}
Black widows (BWs) are a class of eclipsing millisecond pulsars (MSPs) orbiting together with low-mass companions
\citep[$\lesssim 0.05\,\rm M_\odot$;][]{2013IAUS..291..127R, 2013ApJ...775...27C, 2014ApJ...786L...7B}.
It is generally believed that their companions are ablated by
the wind resulting from the $\gamma$-ray illumination and the energetic particles
radiated by MSPs, called the evaporation process
\citep[e.g.][]{1988Natur.334..225K, 1989ApJ...336..507R}.
They are important objects for revealing the evolutionary history of
close neutron star (NS) binary systems,
such as the decay of the surface magnetic field of NSs,
the magnetic braking, the NS equation of state and the accretion efficiency, etc
\citep[e.g.][]{2001ApJ...557..958C, 2012MNRAS.425.1601T, 2019MNRAS.483.5595V, 2021ApJ...908..122G, 2021ApJ...922..158L}.
In addition,
it has been suggested that BWs may be a potential progenitor of isolated MSPs
\citep[e.g.][]{1988Natur.334..227V, 2020MNRAS.495.3656G, 2022MNRAS.515.2725G}.
However,
there is still no agreement on
the origin of BWs
\citep[e.g.][]{2013ApJ...775...27C, 2014ApJ...786L...7B, 2015ApJ...814...74J, 2019ApJ...881...72A, 2022MNRAS.515.2725G}.
	
Since the discovery of the first BW
\citep[PSR B$1957+20$;][]{1988Natur.333..237F},
a large number of BW samples have been detected in the
past decades.
Recently,
a BW pulsar, PSR J1953 + 1844 (i.e. M71E) was discovered by the
Five-hundred-meter Aperture Spherical radio Telescope (FAST)
during the FAST Galactic Plane Pulsar Snapshot survey
\citep[][]{2021RAA....21..107H},
which has a spin period of $4$\,ms.
\citet{cite-key} obtained the orbital period of 53 minutes from the archival data of FAST globular cluster pulsar survey.
The further timing shows that the mass of its companion is $\sim 0.01\,\rm M_\odot$ derived from the mass function of $2.3\times10^{-7}\,\rm M_\odot$ \citep{cite-key}.
If the inclination angle ranges from $25.8^{\circ}$ to $90^{\circ}$,
the companion may be a hydrogen-deficient star,
probably originating from ultracompact X-ray binaries
\citep[UCXBs;][]{zonglin}.
\citet{cite-key} proposed a low mass X-ray binary channel
with an evolved main sequence (MS) donor to explain the formation of PSR J$1953+1844$,
that is, the mass-transfer process starts at the end of the MS phase of the donor.
However,
their model does not include the evaporation process
that plays an important role in the formation of BWs and may widen the orbit.

The increasing number of detected BW samples indicate that
they have two subtypes according to their companion mass ($M_2$):
one is the BWs with $M_2\sim0.01-0.05\,\rm M_\odot$,
and another is the BWs with $M_2\lesssim0.01\,\rm M_\odot$
\citep[see][]{2022MNRAS.515.2725G}.
Previous studies hardly explain the origin of the latter within the Hubble time
\citep[e.g.][]{2013ApJ...775...27C, 2020MNRAS.495.3656G}.
Recently,
\citet{2022MNRAS.515.2725G} further considered the evaporation process
based on the He star donor channel forming UCXBs
\citep[see][]{2021MNRAS.506.4654W}.
They reproduced the characteristics of the BWs
with low-mass companions ($M_2\lesssim0.01\,\rm M_\odot$) and short orbital periods,
such as PSRs J$1719-1438$, J$2322-2650$ and J$1653-0158$, etc.

Accordingly,
the purpose of this paper is to investigate the formation of PSR J$1953+1844$
through the He star donor channel,
in which the binaries can undergo the UCXB phase.
In Section \ref{sec:Numerical methods and assumptions},
we introduce the numerical methods for the binary evolution simulations and the adopted assumptions.
In Section \ref{sec:Results},
we show the results of binary evolution and compare with observations.
Finally,
the relevant discussions and a summary
are given in Sections \ref{sec:Diss} and \ref{sec:Sum}, respectively.
	
\section{Numerical methods and assumptions}\label{sec:Numerical methods and assumptions}
We used the stellar evolution code Modules for Experiments in Stellar Astrophysics (MESA, version 12778; see \citealt{2011ApJS..192....3P, 2013ApJS..208....4P, 2015ApJS..220...15P, 2018ApJS..234...34P, 2019ApJS..243...10P}) to carry out detailed binary evolution computations.
We set the NSs as point masses,
and the initial mass of NS is $M_{\rm NS}^{\rm i}=1.4\,\rm M_\odot$.
%and built the zero-age main-sequence (ZAMS) He stars with mass of $0.32\,\rm M_\odot$.
Meanwhile,
since PSR J$1953+1844$ may locate in globular cluster M$71$,
we constructed zero-age main-sequence He stars
with two initial metallicities ($Z = 0.002, 0.02$).
The initial mass of He star ($M_2^{\rm i}$) and the initial orbital period
are set to be $0.32\,\rm M_\odot$ and
$P_{\rm orb}^{\rm i}=0.013\rm\,d$,
respectively.
In addition,
we adopted the Type-2 opacity tables suitable
for the extra carbon and oxygen because of helium burning
\citep[][]{1996ApJ...464..943I}.
We did not consider the magnetic braking
usually used for the Sun-like star with a radiative core and a convective envelope
\citep[e.g.][]{1983ApJ...275..713R, 2013ApJ...775...27C, 2015ApJS..220...15P, 2022MNRAS.515.2725G}.

During the evolution of the NS+He star systems,
the He star companion fills its Roche lobe as the orbit gradually shrinks
owing to the gravitational wave (GW) radiation.
The orbital angular momentum loss due to the GW radiation
can be calculated by
\citep{1971ctf..book.....L}:
\begin{equation}
	\frac{d J_{\mathrm{GR}}}{d t}=-\frac{32 G^{7 / 2}}{5 c^5} \frac{M_{\mathrm{NS}}^2 M_{\mathrm{2}}^2\left(M_{\mathrm{NS}}+M_{\mathrm{2}}\right)^{1 / 2}}{a^{7 / 2}},
\end{equation}
in which the $G$, $c$ and $a$ are
the gravitational constant, the speed of light in vacuum
and the orbital separation, respectively.
In addition,
we used the `Ritter' scheme to compute the mass-transfer rate
\citep[][]{1988A&A...202...93R}.
We assumed that the fraction of transferred material accreted by NS is $0.5$,
and set the Eddington accretion rate to be $3 \times 10^{-8}\,\rm M_\odot \rm yr^{-1}$
\citep[e.g.][]{2006csxs.book..623T, chen2011A&A, 2021MNRAS.506.4654W,2023MNRAS.526..932G}.

\citet{2008A&A...486..523L} suggested that for the irradiated pure helium discs,
the thermal-viscous instability of accretion discs could be triggered
if the mass-transfer rate is lower than $\sim5 \times 10^{-10}\,\rm M_\odot \rm yr^{-1}$.
At this moment,
we assumed that the He star companion starts to be evaporated by the pulsar radiation.
We used a simple prescription proposed by \citet{1992MNRAS.254P..19S}
to calculate the mass-loss rate of donor caused by the evaporation process:
\begin{equation}
	\dot{M}_{\rm 2, evap}=-\frac{f}{2 \nu_{\rm2, esc }^2} L_{\mathrm{P}}\left(\frac{R_2}{a}\right)^2,
\end{equation}
where $f$, $\nu_{\rm2, esc }$ and $R_2$ are
the evaporation efficiency, the escape velocity at the donor surface
and the donor radius, respectively.
In this work,
we calculated the binary evolution with different evaporation efficiencies,
i.e. $f = 0, 0.01, 0.1$.
$L_{\rm P} = 4\pi^2I\dot P_{\rm spin}/P_{\rm spin}^3$
is the spin-down luminosity of pulsars,
in which $I=10^{45}\rm\,g\,cm^2$ is the pulsar moment of inertia,
$P_{\rm spin}$ and $\dot P_{\rm spin}$
are the spin period of pulsars and its derivative, respectively.
We set the initial value of $P_{\rm spin}$ and $\dot P_{\rm spin}$
to be $3$\,ms and $1.0\times10^{-20}\rm\,s\,s^{-1}$,
and used a constant braking index (i.e. $n=3$)
to calculate the evolution of $L_{\rm P}$
\citep[e.g.][]{2013ApJ...775...27C, 2015ApJ...814...74J}.

%\begin{equation}
%	-\dot{M}_2=\dot{M}_0 \exp \left[\frac{R_2-R_{\mathrm{L}, 2}}{H_{\mathrm{P}} / \gamma(q)}\right],
%\end{equation}

%\begin{equation}
%	\frac{R_{\mathrm{L}, 2}}{a}=\frac{0.49 q^{-2 / 3}}{0.6 q^{-2 / 3}+\ln \left(1+q^{-1 / 3}\right)},
%\end{equation}
\begin{figure*}
		\centering\includegraphics[width=\columnwidth*11/6]{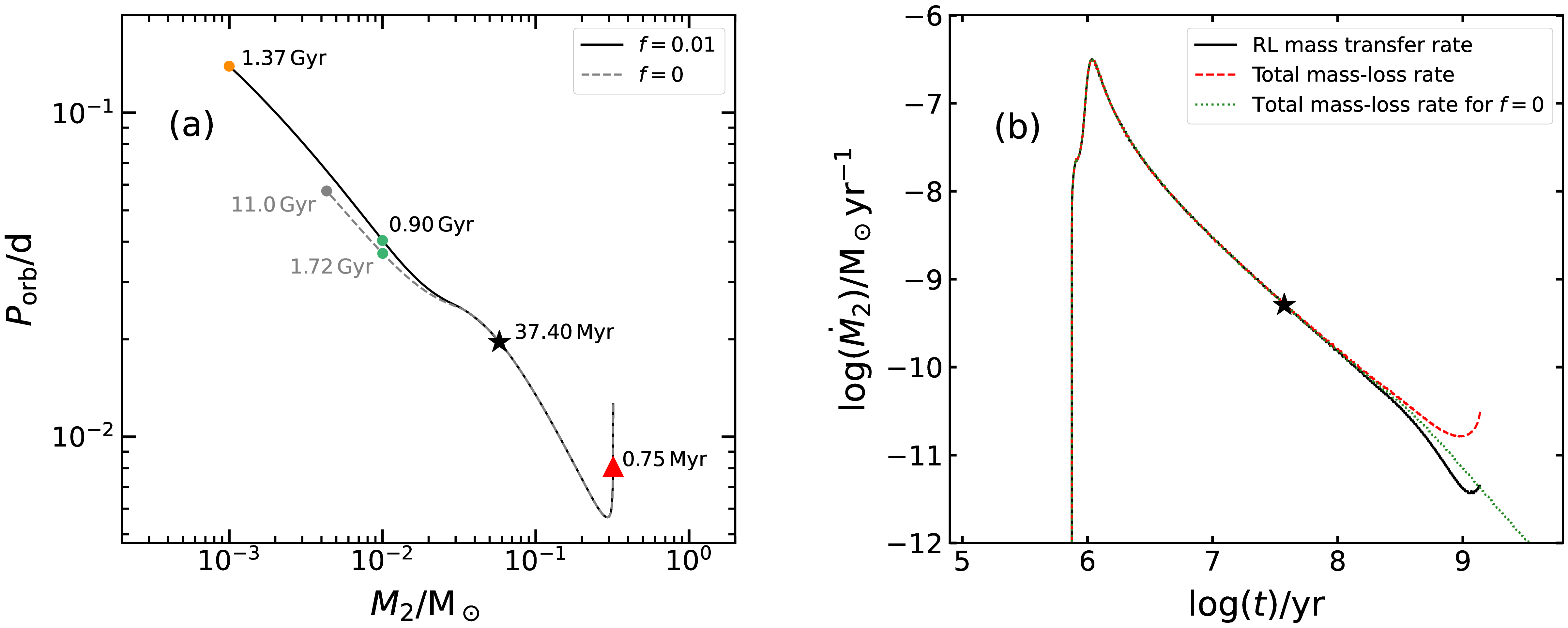}
		\caption{A typical example for the evolution of a NS+He star system that can undergo UCXB phase,
			in which $M_{\rm NS}^{\rm i}=1.4\,\rm M_\odot$, $M_2^{\rm i}=0.32\,\rm M_\odot$,
			$P_{\rm orb}^{\rm i}=0.013\rm\,d$, $Z=0.02$ and $f=0.01$.
			Panel (a): Evolutionary track of NS+He star system in the orbital period versus donor mass diagram.
			The red triangle and the black star indicate the moments
			when the mass transfer and the evaporation process start, respectively.
			%The beginning of evaporation process is marked with the black stars.
			The green and orange dots denote the moments when the donor mass decreases to $0.01\,\rm M_\odot$ and $0.001\,\rm M_\odot$, respectively.
			The gray dashed line denotes the evolutionary track of binary
			without evaporation process,
			and the code stops at $t\sim11.0$\,Gyr (the gray dot) because of hitting the equation-of-state limits
			\citep[][]{2021MNRAS.506.4654W}.
			Panel (b): Evolution of mass-transfer rate (black line) and total mass-loss rate (red dashed line) as a function of time.
			The green dotted line denotes the evolutionary track of mass-loss rate
			for the model without evaporation process.}
		\label{fig:M71-pm}
\end{figure*}
\section{Results}\label{sec:Results}
\subsection{A typical example for binary evolution}\label{sec:example}
Fig.\,\ref{fig:M71-pm} shows a representative example of the evolution of a NS+He star system that can experience UCXB phase,
in which we set the evaporation efficiency to be a typical value of $0.01$.
In the early stage of binary evolution,
the rapid shrinking of the orbital separation is dominated by the loss of angular momentum owing to the GW radiation.
At $t\sim0.75$\,Myr,
the He star companion begins to fill its Roche lobe and the orbital period drops to $0.2$\,h,
indicating that this binary system starts to appear as an UCXB (see Fig.\,\ref{fig:M71-pm}a).
At $t\sim1.05$\,Myr,
the binary evolves to the minimum orbital period ($\sim0.14$\,h)
and the companion star decreases its mass to $\sim0.30\,\rm M_\odot$.
After that,
the helium burning in the center of He star starts to fade
and the He star gradually reaches a mildly degenerate state,
indicating that this donor has a negative mass-radius exponent
\citep[see][]{2021MNRAS.506.4654W}.

At $t\sim37.40$\,Myr,
the mass-transfer rate decreases to $\sim5\times10^{-10}\,\rm M_\odot\rm\,yr^{-1}$,
thus triggering the thermal-viscous instability of accretion discs.
At this moment,
the donor mass and the orbital period is
$\sim0.058\,\rm M_\odot$ and $\sim28.26$ minutes, respectively.
Meanwhile,
the NS has accreted $\sim0.1\,\rm M_\odot$ of material
and spins up to become a MSP,
resulting in that the donor begins to be evaporated by the pulsar radiation.
From Fig.\,\ref{fig:M71-pm},
we can see that the evaporation process increases the mass-loss rate of the donor,
which could widen the orbital separation faster
\citep[e.g.][]{2013ApJ...775...27C, 2022MNRAS.515.2725G}.
The mass-transfer rate starts to increase at $t\sim1$\,Gyr,
resulting from more rapid expansion of the degenerate He donor
caused by the effective mass loss.
In addition,
we note that the mass of the donor can decrease to $<10^{-2}\,\rm M_\odot$
or even $<10^{-3}\,\rm M_\odot$ within the Hubble time.
	
\subsection{Comparison with observations}\label{sec:observation}
\begin{figure*}
	\centering\includegraphics[width=\columnwidth*10/6]{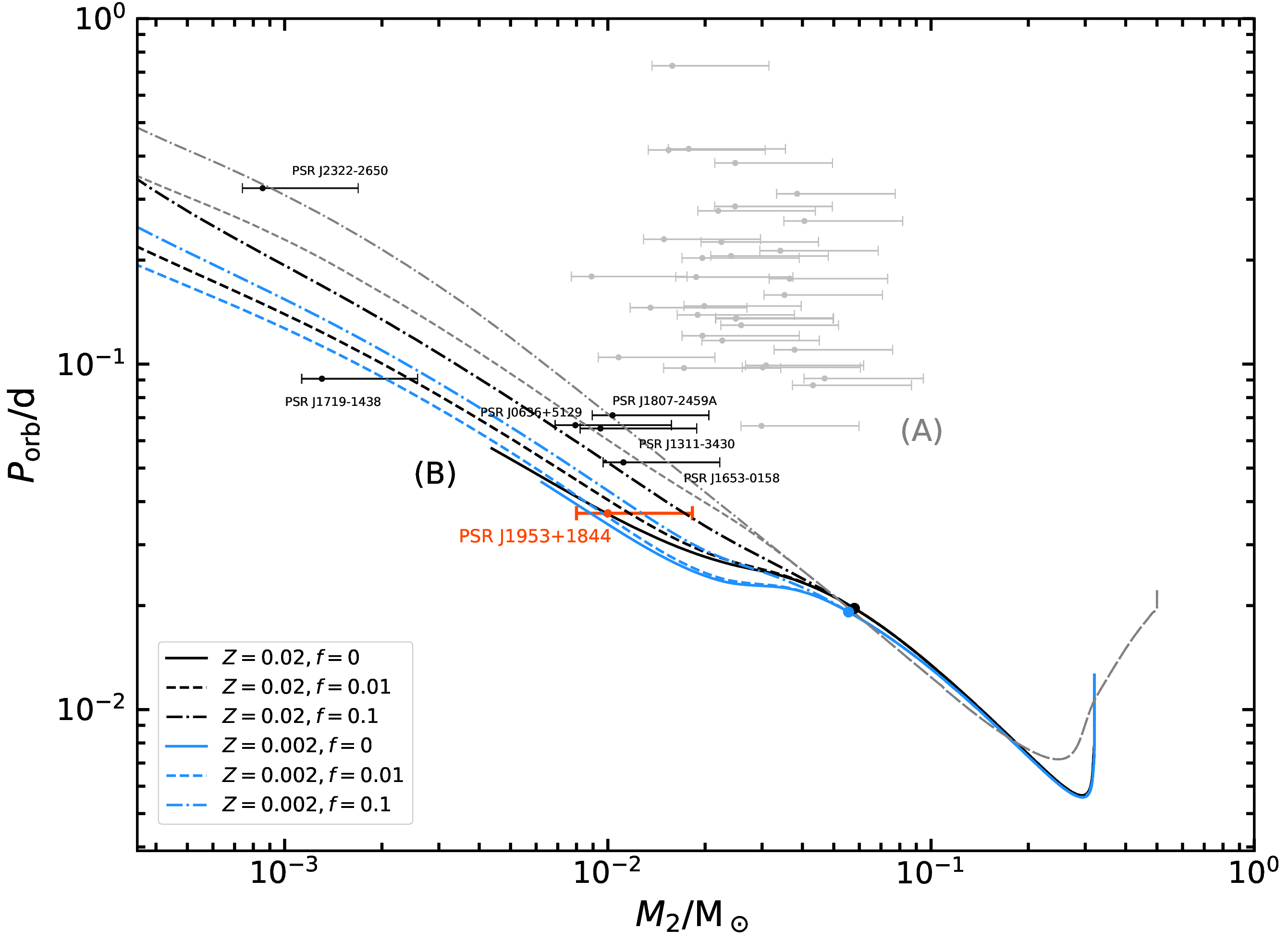}
	\caption{The evolutionary tracks of NS+He star systems with different $f$-values and  metallicities (the black and blue lines)
in the orbital period versus donor mass diagram,
			in which $M_{\rm NS}^{\rm i}=1.4\,\rm M_\odot$, $M_2^{\rm i}=0.32\,\rm M_\odot$ and $P_{\rm orb}^{\rm i}=0.013\rm\,d$.
			The blue and black dots indicate the moment	when the evaporation process starts.
The gray dashed and dashed–dotted lines represent
the binary models with
($M_2, P_{\rm orb}^{\rm i}(\rm d),$ $f$) = ($0.5\,\rm M_\odot, 0.022, 0.01$)
and ($M_2, P_{\rm orb}^{\rm i}(\rm d),$ $f$) = ($0.5\,\rm M_\odot, 0.022, 0.1$),
in which the simulated data are taken from \citet{2022MNRAS.515.2725G}.
			The red sample denotes PSR J$1953+1844$.
The gray samples in region (A) are the BWs with $M_2\sim0.01-0.05\,\rm M_\odot$,
			and the black samples in region (B) are the BWs that can be explained by the He star donor channel.
			The left and right sides of the error bars correspond to the companion mass at orbital inclination angle of $90^{\circ}$ and $26^{\circ}$ (the $90\%$
			probability limit), respectively.
			The observed data are taken from the ATNF Pulsar Catalogue, \href{http://www.atnf.csiro.au/research/pulsar/psrcat}{http://www.atnf.csiro.au/research/pulsar/psrcat}
			\citep[version 1.70, 2023 May;][]{2005AJ....129.1993M}.
		}
		\label{fig:M71E-ob}
\end{figure*}

We performed a series of complete binary evolution computations
to reproduce the characteristics of PSR J$1953+1844$.
Fig.\,\ref{fig:M71E-ob} represents the evolutionary tracks of NS+He star systems
with different $f$-values in the orbital period versus donor mass diagram.
Meanwhile,
PSR J$1953+1844$ may be located in globular cluster M$71$
\citep[][]{cite-key},
corresponding to a metallicity of $0.002$
\citep[][]{1996AJ....112.1487H}.
Thus,
we also calculated the binary evolution
with an initial metallicity of $0.002$ for the He star companions.
We note that for a given evaporation efficiency,
the binaries have shorter orbital periods in lower metallicity.
This is because lower metallicity leads to
a smaller donor radius and mass-transfer rate,
resulting in that the orbit expand more slowly
\citep[see, e.g.][]{2010A&A...515A..88W}.
In addition,
it is worth noting that
our channel not only produces the BWs with low-mass companions
\citep[$M_2\lesssim0.01\,\rm M_\odot$;][]{2022MNRAS.515.2725G},
but also explains the BWs with ultra-compact orbits.
The minimum orbital period of BWs
that this channel can produce is $\sim28$ minutes.

From Fig.\,\ref{fig:M71E-ob},
we can see that our simulations can explain
the short orbital period and the low companion mass of PSR J$1953+1844$,
indicating that this BW may originate from the He star donor channel.
Meanwhile,
our simulations show that
PSR J$1953+1844$ may be produced by a evaporation efficiency of $\lesssim0.1$.
In addition,
it is worth noting that
PSR J$1953+1844$ does not show the eclipsing events in the observations,
and \citet[][]{cite-key} speculated that its orbit is not edge-on.
Nevertheless,
this channel can still produce the characteristics of PSR J$1953+1844$
even without evaporation.
Therefore,
the detected PSR J$1953+1844$ provides a strong support
for the He star donor channel.

\section{Discussions}\label{sec:Diss}
\subsection{Gravitational wave signals}
The detection of high-frequency GW signals from
double black hole and double NS coalescence events
open a new era of multimessenger astrophysics
\citep[GW150914 and GW170817;][]{2016PhRvL.116f1102A, 2017PhRvL.119n1101A}.
On the other hand,
the future space-based GW observatories
(i.e. Laser Interferometer Space Antenna (LISA), TaiJi and TianQin) can be used to
detect the low-frequency GW signals radiated by the inspiral processes of close binaries,
such as UCXBs, AM CVn and double white dwarfs (WDs), etc
\citep[e.g.][]{2010NewAR..54...87N, 2020ApJ...900L...8C, 2021MNRAS.506.4654W, 2022ApJ...935....9C, 2023MNRAS.522.5358F}.
The recent detection of the BWs with short orbital periods
(e.g. PSRs J$1653-0158$ and J$1953 + 1844$)
indicate that the BWs may also be the potential sources of
low-frequency GW.
Thus,
in this subsection,
we discuss the possibility of BWs formed by this channel
being detected by space-based GW observatories.

By considering that the GW signal is monochromatic,
the characteristic strain amplitude of UCXBs based on $4$\,yr of
LISA observations can be expressed as
\citep[e.g.][]{2018PhRvL.121m1105T, 2020ApJ...896..129C}:
\begin{equation}
		h_{\mathrm{c}} \approx 2.5 \times 10^{-20}\left(\frac{M_{\text {chirp }}}{\mathrm{M}_{\odot}}\right)^{5 / 3}\left(\frac{f_{\mathrm{GW}}}{\mathrm{mHz}}\right)^{7 / 6}\left(\frac{15\, \mathrm{kpc}}{d}\right),
\end{equation}
in which $f_{\rm GW}=2/P_{\rm orb}$ is the GW frequency,
and $d$ is the distance from the source to the detector.
The chirp mass can be  written as
\citep{2018PhRvL.121m1105T}:
\begin{equation}
	M_{\text {chirp }}=\frac{c^3}{G}\left(\frac{5 \pi^{-8 / 3}}{96} f_{\mathrm{GW}}^{-11 / 3} \dot{f}_{\mathrm{GW}}\right)^{3 / 5},
\end{equation}
where $\dot{f}_{\mathrm{GW}}$ is the GW frequency derivative.

\begin{figure*}
		\centering\includegraphics[width=\columnwidth*10/6]{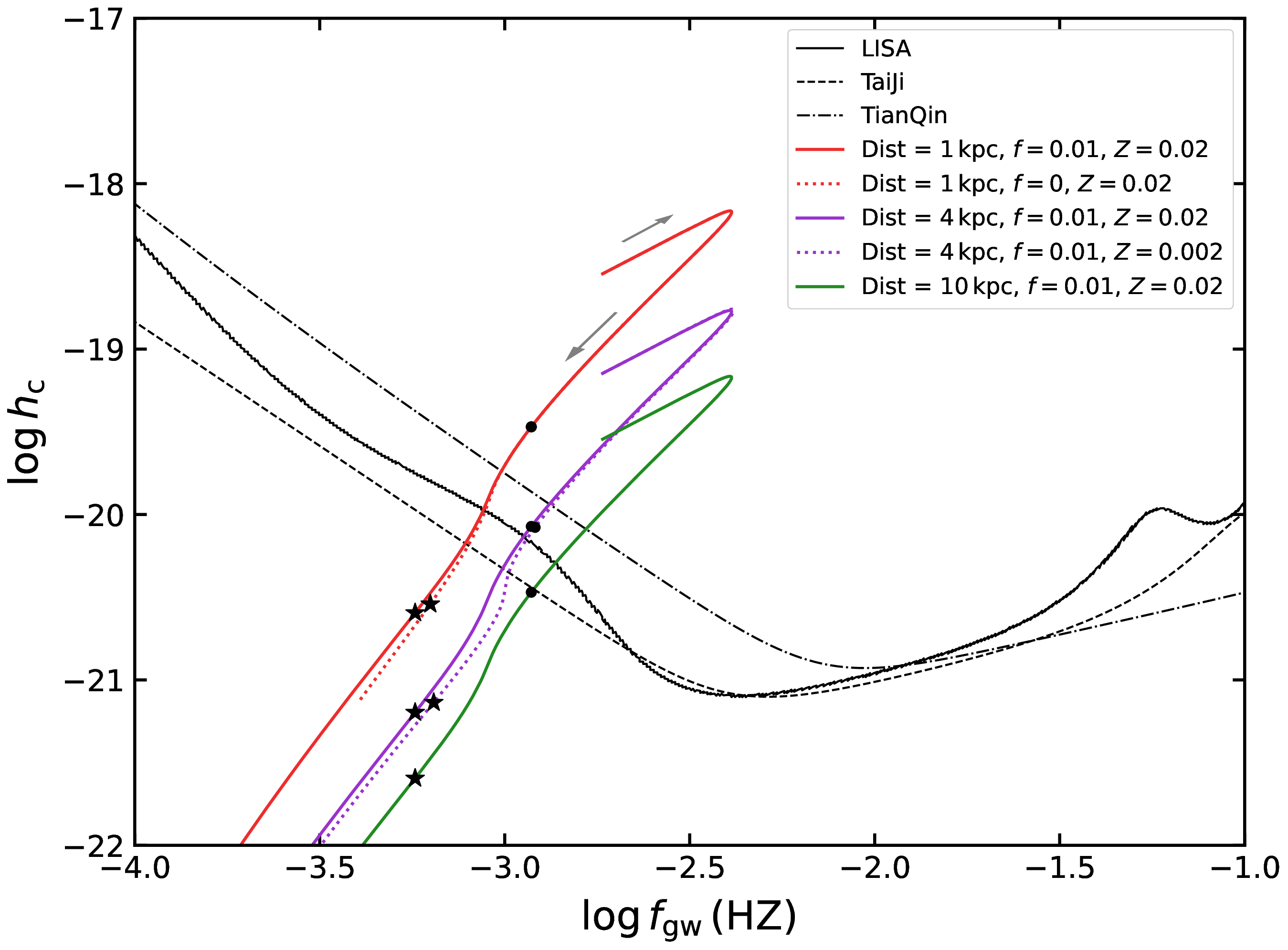}
		\caption{Evolutionary tracks of the NS+He star systems with different evaporation efficiencies, metallicities and the distances from the source to the detectors
			in the characteristic strain amplitude vs. GW frequency diagram,
			where $M_{\rm NS}^{\rm i}=1.4\,\rm M_\odot$, $M_2^{\rm i}=0.32\,\rm M_\odot$ and
			$P_{\rm orb}^{\rm i}=0.013\rm\,d$.
			The evolutionary tracks follow the direction of the gray arrows.
			The black dots and stars indicate the moments when the evaporation process starts and the companion mass decreases to $0.01\,\rm M_\odot$, respectively.
			The black solid,  dashed and dashed–dotted lines represent
			the sensitive curves for LISA, TaiJi and TianQin, respectively.
		}
		\label{fig:GW}
\end{figure*}

Fig.\,\ref{fig:GW} shows the characteristic GW strain of the NS+He star binaries
as a function of GW frequency,
as well as the sensitive curves for LISA, TaiJi and TianQin
\citep[e.g.][]{2019CQGra..36j5011R, 2019PhRvD.100d3003W, 2020NatAs...4..108R}.
We can see that metallicity and evaporation efficiency
have no significant effect on the characteristic GW strains.
In addition,
the BWs originating from this channel cannot be detected
by GW observatories if the distance exceeds $10$\,kpc.
If we assume that PSR J$1953 + 1844$ is located in M71
(i.e. the distance is $4$\,kpc),
then it cannot be visible
as a low-frequency GW source (see the purple solid line).
Table\,\ref{table:1} shows the characteristics of binaries
(i.e. the minimum donor mass and maximum orbital period)
that can be detected by different detectors.
Meanwhile,
the parameter of the binaries at the beginning of evaporation is
($M_2, P_{\rm orb}$) = ($0.058\,\rm M_\odot, 28.26$ minutes).
Accordingly,
although the parameter space of the BWs
that can be visible as GW sources is narrow,
we still expect that
the future space-based GW observatories will be helpful to detect
the BWs with extremely short orbital periods.

\begin{table}
		\centering
		\caption{
			Characteristics of binaries that can be detected by different detectors.
			Dist is the distances from the source to the detectors;
			$M_2^{\rm min}$ and $P_{\rm orb}^{\rm max}$ are the
			minimum companion mass and the maximum orbital period
			of the binaries that can be detected by the detectors, respectively.
		}
		\label{table:1}
		\begin{tabular}{ c  c c c cccc c c c cc c }
			\toprule
			\hline 
			Detectors&& Dist &$M_2^{\rm min}$ &$P_{\rm orb}^{\rm max}$\\ %&$\Delta t$\\
			&&(kpc)&$(\rm M_\odot)$   &(minutes)\\%&\\
			\hline 
			LISA		  && $1$&$0.026$ &$38$\\%&\\
			&& $4$& $0.053$&$30$\\%&\\
			\hline 
			TaiJi		   && $1$&$0.018$ &$43$\\%&\\
			&& $4$&$0.040$&$34$\\%&\\
			\hline 
			TianQin    && $1$& $0.039$&$34$\\%&\\
			&& $4$& $...$&$...$\\%&\\
			\hline
		\end{tabular}
\end{table} 
	
\subsection{Isolated MSP}
At present,
more than two hundred isolated MSPs have been detected.
It is generally believed that
BWs may be one of the progenitors of isolated MSPs,
although the evolution process is still unclear
\citep[e.g.][]{1988Natur.334..227V, 2020MNRAS.495.3656G, 2022MNRAS.515.2725G}.
Especially,
two pulsars with planetary-mass companion ($M_2\sim10^{-3}\,\rm M_\odot$)
have been detected,
i.e. PSR J$1719-1438$ \citep[][]{2011Sci...333.1717B}
and PSR J$2322-2650$ \citep[][]{2018MNRAS.475..469S},
which may be two bridging objects between BWs and isolated MSPs.
However,
previous studies of evaporating the MS star with pulsar radiation
is difficult to explain the formation of PSRs J$1719-1438$, J$2322-2650$
and the isolated MSPs within the Hubble time
\citep[e.g.][]{2013ApJ...775...27C, 2020MNRAS.495.3656G}.

\citet{1985ApJ...289..244R} suggested that
He degenerate companion may experience the tidal disruption
if its mass decreases to $\sim10^{-3}\,\rm M_\odot$\footnote{
By approximately treating the donor as a zero-temperature and free degenerate gas,
\citet{1985ApJ...289..244R} assumed that
NS can tidally disrupt its He degenerate donor
if
%($\rm d$\,log$R_{\rm L}$)/($\rm d$\,log$M_2$)
$\frac{d\rm\,log\textit{R}_{\rm L}}{d\rm\,log\textit{M}_2}\textgreater\frac{d\rm\,log\textit{R}_{\rm 2}}{d\rm\,log\textit{M}_2}\simeq-\frac{1}{3}$,
in which $R_{\rm L}$ is the Roche lobe radius of the donor
related to the donor mass, NS mass, and orbital separation
\citep[see also][]{1992MNRAS.254P..19S}.}.
Meanwhile,
\citet{2022MNRAS.515.2725G} found that
the companion mass can decrease to $<10^{-3}\,\rm M_\odot$ within the Hubble time
for the He star donor channel,
indicating that this channel is the potential progenitor of isolated MSPs.
In this work,
if we assume that the initial metallicity of the companion is $0.002$
(i.e. PSR J$1953+1844$ is located in M71) and the evaporation efficiency is $0.01$,
then the companion mass will decrease from $0.01\,\rm M_\odot$ to $\sim10^{-3}\,\rm M_\odot$ after $\sim0.50$\,Gyr.
Accordingly,
we expect that PSR J$1953+1844$ will eventually evolve into an isolated MSP.

\subsection{Comparison to previous studies}
It is still under highly debate for the companion nature of PSR J$1953+1844$.
By assuming that the companion is a brown dwarf,
\citet{cite-key} concluded that
the companion mass may range from $0.047\,\rm M_\odot$ to $0.097\,\rm M_\odot$
according to the mass-radius relationship.
However,
this mass range requires a narrow range of
orbital inclination angle from $3.8^{\circ}$ to $12.1^{\circ}$.
On the other hand,
the companion mass
ranges from $8\times10^{-3}\,\rm M_\odot$ to $15\times10^{-3}\,\rm M_\odot$
based on its low mass function of $2.3\times10^{-7}\,\rm M_\odot$,
corresponding to a larger range of the
orbital inclination angle from $25.8^{\circ}$ to $90^{\circ}$
\citep[e.g.][]{cite-key, zonglin}.
Thus,
the companion of PSR J$1953+1844$
is more likely to have a low mass of $\sim0.01\,\rm M_\odot$.
In addition,
the simulations of \citet{cite-key} show that
for the evolved MS donor channel,
the binaries could undergo the UCXB phase,
and the companion mass decreases to $\sim0.01\,\rm M_\odot$
at $t\sim10^{10}$\,yr.
Meanwhile,
they claimed that PSR J$1953+1844$ may be the descendant of redback.
\citet{2023MNRAS.525.2708C} recently studied the evolved MS donor channel
with a suitable magnetic braking model,
and they found that orbital periods shorter than $1$\,h require significant fine-tuning.

In this work,
the companions can decrease their masses to $\sim0.01\,\rm M_\odot$ at $t\sim10^{9}$\,yr,
and this time-scale is much shorter than that in \citet{cite-key}.
Meanwhile,
our simulations suggest that
the progenitor of PSR J$1953+1844$ is an UCXB instead of a redback.
In addition,
the minimum orbital period of BWs in their simulations is $\sim38.68$ minutes,
which is slightly higher than our results ($\sim28$ minutes).
Moreover,
the minimum initial mass of the He star companions
in this channel is $0.32\,\rm M_\odot$,
which is the lower mass limit for He stars to ignite the central helium
\citep{2002MNRAS.336..449H}.
Thus,
the surface of donors in our simulations shows the enrichment of the metal elements
(e.g. C and O) after the evaporation process starts owing to the helium burning,
and the abundance of metal elements increases with the initial mass of the He star donor.
On the other hand,
the donor ($M_2\approx0.01\,\rm M_\odot$) in \citet{cite-key} has no metal enrichment,
and is mainly composed of helium and hydrogen
because the donor does not undergo the central helium burn
(Chen H.-L. 2023, private communication).
Furthermore,
\citet{zonglin} derived the evolutionary tracks of UCXBs
to explain the properties of PSR J$1953+1844$ by using the analytical approach,
and they suggested that its companion can be a CO WD or a He WD.
More massive He star donor is needed if the companion of PSR J$1953+1844$ is a CO WD.
It is expected that the future spectroscopic studies
will provide more detailed properties for the companion of PSR J$1953+1844$.

\section{Summary}\label{sec:Sum}
PSR J$1953+1844$ has the shortest orbital period ($\sim53$ minutes)
among the detected BWs,
and its companion mass is $\sim0.01\,\rm M_\odot$ derived by its mass function.
Using the stellar evolution code MESA,
we explored the origin of PSR J$1953+1844$
by considering the evaporation process
based on the He star donor channel,
where the binaries could experience the UCXB phase.
Our simulations reproduce the characteristics of PSR J$1953+1844$,
i.e. the low companion mass
and the short orbital period.
Thus,
this work provides an alternative way for the origin of PSR J$1953+1844$,
and implies that this object may originate from an UCXB.
%This work suggests that
%PSR J$1953+1844$ originates from UCXBs rather than redbacks.
Meanwhile,
we note that for the He star donor channel,
the minimum orbital period of the formed BWs is $\sim28$ minutes,
corresponding to the companion mass of $0.058\,\rm M_\odot$.
We then discussed whether the BWs formed by this channel
can be detected by the future space-based GW observatories
like LISA, TaiJi and TianQin.
We found that the GW observatories
may be helpful for the detection of BWs with extremely short orbital periods.
In addition,
PSR J$1953+1844$ may evolve into an isolated MSP eventually,
and this object provides a support for the He star donor channel.

\section*{Acknowledgements}
We acknowledge the anonymous referee for the valuable comments
that help to improve this paper.
We thank Prof. Jinlin Han, Prof. Hailiang Chen, Zonglin Yang and Sivan Ginzburg
for useful discussions and comments.
This study is supported by the
National Natural Science Foundation of China (Nos 12041301, 12121003 and 12225304), the National Key R\&D Program of China (Nos 2021YFA0718500 and 2021YFA1600404), the Western Light Project of CAS (No. XBZG-ZDSYS-202117), the science research grant from the China Manned Space Project (No. CMS-CSST-2021-A12), the Yunnan Fundamental Research Project (No. 202201BC070003),  and the Frontier Scientific Research Program of Deep Space Exploration Laboratory (No. 2022-QYKYJH-ZYTS-016).

\section*{Data availability}
Results will be shared on reasonable request to corresponding author.

\bibliographystyle{mnras} % style aa.bst
\bibliography{1bib.bib}					
%%%%%%%%%%%%%%%%%%%%%%%%%%%%%%%%%%%%%%%%%%%%%%%%%%

%%%%%%%%%%%%%%%%% APPENDICES %%%%%%%%%%%%%%%%%%%%%

%%%%%%%%%%%%%%%%%%%%%%%%%%%%%%%%%%%%%%%%%%%%%%%%%%

% Don't change these lines
%\bsp	% typesetting comment
\label{lastpage}
\end{document}